# CSRE4SOC (CSR evaluation for software companies)

Elisa Jiménez
Institute of Technology and Informations Systems
University of Castilla-La Mancha
Ciudad Real, Spain
elisa.jimenez@uclm.es

Coral Calero
Institute of Technology and Informations Systems
University of Castilla-La Mancha
Ciudad Real, Spain
coral.calero@uclm.es

Mª Ángeles Moraga
Institute of Technology and Informations Systems
University of Castilla-La Mancha
Ciudad Real, Spain
mariaangeles.moraga@uclm.es

*Abstract*— Software development companies are increasingly concerned about their impact on the environment. This is translated into the incorporation of actions related to software sustainability in their Corporate Social Responsibility (CSR) document.

CSR reflects a company's obligations to society and the environment. However, we have found that companies do not always have the necessary knowledge to be able to include actions related to software sustainability. Moreover, there is still a lot of work to be done, as the number of actions they incorporate is often insufficient.

Taking all this into account, we consider it essential for software development companies to have a tool that allows them to assess their level of software sustainability, based on the actions of their CSR, and to automatically provide them with a series of improvements to advance their level of software sustainability.

Therefore, this paper introduces CSRE4SOC, a tool for the evaluation and monitoring of the software sustainability level of software development companies according to their CSR.

## I. TECHNICAL DESCRIPTION

The objective of social responsibility is to contribute to sustainable development [1]. CSR involves the voluntary integration by companies of social and environmental concerns in their business operations and relationships with their partners [2]. In general, the CSR document identifies five dimensions: voluntariness, economic, social, stakeholders and environmental.

However, the incorporation of software sustainability aspects to meet the environmental dimension has been relegated to the background, as the negative impact of the development and use of software on the environment has not traditionally been considered by society.

Fortunately, this perception is changing, and software development companies are becoming increasingly aware of the importance of software sustainability.

According to [3], software sustainability is about the capability of software to last a long time by using only the resources that are strictly needed.

In [4] the authors identify three kinds of resources needed by software life cycle processes that can be used to directly obtain the dimensions of software sustainability: human, economic and environmental.

Software companies should focus their efforts on improving these three dimensions, which is why they need to incorporate actions related to each of the dimensions in their CSR.

In [5], the authors identify a set of actions for each of the dimensions of software sustainability that software development companies should incorporate into their CSR. They also establish how to calculate a company's level of software sustainability based on the actions defined in its CSR.

Using this research as a starting point, the CSRE4SOC tool has been developed, which allows the assessment of compliance with the actions of each of the three dimensions of sustainability included in the CSR of a software development company in order to determine its level of sustainability.

In addition, the tool proposes a series of improvement actions for the company to carry out in order to increase its level of software sustainability (e.g., "A process to collect the software energy consumption should be defined." or "The KW/h required by each software functionality should be reduced."). Once the company has implemented the actions it deems appropriate, it can re-evaluate itself and check whether it has really improved. In order to visualize the evolution of the levels of software sustainability achieved by the company, the tool keeps a history of results and provides a graphical display of this evolution.

CSRE4SOC has been developed as a web application using the model-view-controller (MVC) architecture type. Both the model and the controller have been developed mainly using the Java language with the help of some technologies such as Spring or Maven. As for the frontend, JavaScript has been used to communicate with the controller, and tagging languages such as HTML with Bootstrap for the visual design of the user interface.

The main functionalities of the CSRE4SOC tool are explained in detail below:

- **Analysis of actions:** CSRE4SOC provides a form showing the total set of actions for each of the software sustainability dimensions, where companies can conveniently choose the ones they have implemented in their CSR document. Fig. 1 shows an example of the form in which companies can choose the actions they have implemented in their CSR (for the environmental dimension).

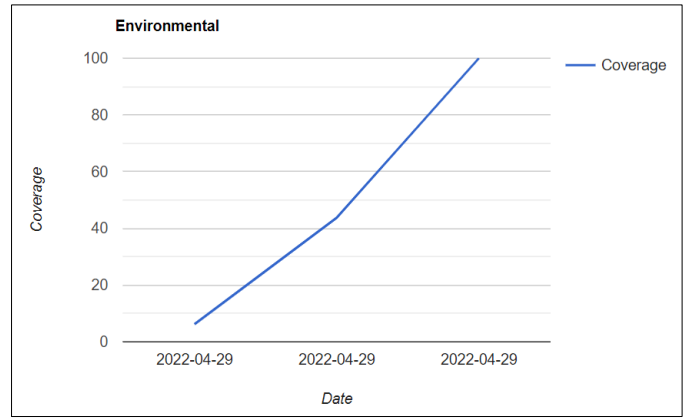

Fig. 1 Questionary

- **Calculation of the sustainability level**: Once the actions taken into account by the company's CSR are indicated, the CSRE4SOC tool automatically calculates the software sustainability level. To do so, the steps proposed in [6] are followed.

Finally, CSRE4SOC provides a graphical display of the results. Fig. 2 shows an example of the sustainability levels obtained after an analysis.

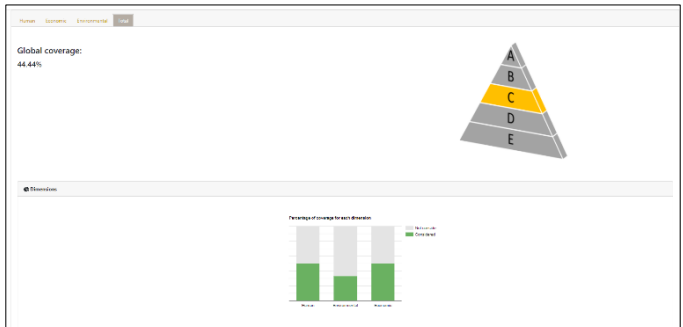

Fig. 2 Results

- **Follow-up and improvement:** In addition to the information resulting from the calculations, this tool provides a series of recommendations for companies to implement and achieve more sustainability actions and thus reach a higher level of software sustainability.

- **Evolution:** Finally, CSRE4SOC offers a section in which companies can track their analysis over time as well as the levels obtained in each of the dimensions. Fig. 3 shows the section of the tool that represents the evolution of the sustainability levels that a company has achieved over time.

Fig. 3 Evolution

## II. RELEVANCE AND NOVELTY

Nowadays, the Information and Communications Technology (ICT) sector is growing rapidly as society becomes increasingly dependent on software.

However, we are often unaware of the impact this can have on the environment as the use and development of software has an impact thereon. It is therefore essential that organisations and professionals involved in software development are aware of the environmental impact of software development and can implement certain practices to reduce environmental damage by offering their users more efficient products.

Companies are therefore required to incorporate actions related to software sustainability in their CSR, and to analyse their CSR documents from time to time in order to know their current status and improve it.

However, this work can be very tedious for companies as it is a time-consuming task to perform by hand, so automating this process would be of great help to them. There is nowadays no tool that allows companies to automatically evaluate their level of sustainability based on the actions defined in their CSR. The current deficiencies in the instruments for sustainability analysis in companies have marked the main objective of this work, which consists of developing a scorecard for the evaluation and monitoring of compliance with the actions of each of the three dimensions of sustainability included in the CSR of a software development company in order to determine its level of software sustainability.

REFERENCES

[1] American National Standards Institute (ANSI) Publications, *ISO26000*. Switzerland., 2010.
[2] European commision, *Green Book*. Washington DC: Department of the Treasury. Financial Management Service., 2000.
[3] M. Dick y S. Naumann, «Enhancing Software Engineering Processes towards Sustainable Software Product Design. », 2010, pp. 706-715.
[4] C. Calero y M. Piattini, «Puzzling out software sustainability», *Sustainable Computing: Informatics and Systems*, vol. 16, pp. 117-124, 2017.
[5] C. Calero, M. Á. Moraga, y M. Piattini, *software sustainability*. Springer. Available at: https://link.springer.com/book/10.1007/978-3-030-69970-3
[6] S. Abrhäo y C. Calero, *Calidad y sostenibilidad de sistemas de información en la práctica.* España: Rama, 2022.